\documentclass{article}
\usepackage[utf8]{inputenc}
\usepackage{tikz}
\usepackage{graphicx}
\usetikzlibrary{positioning}
\usepackage[T1]{fontenc}
\usepackage{amsmath}

\usepackage{mathpazo}
\usepackage[english]{babel}
\usepackage{authblk}
\usepackage{setspace}
\usepackage[margin=1.25in]{geometry}
\usepackage{graphicx}
\graphicspath{ {./figures/} }
\usepackage{subcaption}
\usepackage{subfiles}
\usepackage{braket}
\usepackage[style=ieee, 
citestyle=numeric-comp,
sorting=none]{biblatex}
\DeclareUnicodeCharacter{2212}{-}
\title{Quantum State Transfer Between NV center - $13_C$ 
System Coupled To A CPW Cavity}

\author[1]{Soubhik Pal}
\author[1]{Chiranjib Mitra}

\affil[1]{Department of Physics, Indian Institute of Science Education and Research, Kolkata, India.}

\date{}

\begin{document}

\maketitle

\begin{abstract}
Quantum state transfer is a very important process in building a quantum network when information from flying Qubit is transferred to the stationary Qubit in a node via a quantum state transfer. NV centers due to their  long coherence time and the presence of nearby $13_C$ nuclear spin is an excellent candidate for multi Qubit quantum memory. Here we propose a theoretical description for such a quantum state transfer from a cavity to a nearest neighbour $13_C$ nuclear spin of a single Nitrogen vacancy center in diamond; it shows great potential in realizing scalable quantum networks and quantum simulation. The full Hamiltonian was considered with the zeroth order and interaction terms in the Hamiltonian and the theory of effective hamiltonian theory was applied. We study the time evolution of the combined cavity-$13_C$
    state through analytical calculation and simulation using QuTip. Graphs for state transfer and fidelity measurement are presented here. We show that our theoretical description verifies a high fidelity quantum state transfer from cavity to $13_C$ center by choosing suitable system parameters. 
\end{abstract}


\section{Introduction}
With the remarkable progress in Quantum Information processing and quantum computation, a scalable Quantum Network is no longer just a dream but a reality. An important challenge for an effective and efficient quantum network is the possibility of transferring information through various devices. Quantum State Transfer from a flying qubit to a stationary qubit with good fidelity in a node thus, is a necessity for short distance communication of information and scalability \cite{1} in quantum nodes between cavities and internal state of atom/ion as well as entanglement distribution \cite{2}. QST has been shown and attempted for several different systems like superconducting-optomechanical systems\cite{3} and other systems like Josephson qubits using a resonant cavity\cite{4} and also discussed in other publications \cite{5,6,7,8,9}. It has been demonstrated in Nitrogen Vacancy center qubits as well \cite{10}; recently QST was studied between Coplanar waveguide  cavities and the vibrational modes of micromechanical cantilever \cite{11}. Nitrogen Vacancy(NV) centers are shown to be a very good candidate for quantum memories for their long coherence times even at room temperatures \cite{12} and for their very promising spin dependent optical properties \cite{13}. They have the inherent advantage of using adjacent $13_C$ states which can be used as qubits along with the NV centers \cite{14} and can also be used for entanglement swapping as well \cite{15}. These $13_C$ centers are shown to have high coherence times themselves \cite{16}. NV centers are shown to have good coupling in CPW resonators \cite{17}. In this article we propose a theoretical description of state transfer directly from a CPW cavity to a NV-$13_C$ system coupled to it. We consider the full Hamiltonian including the interaction terms between NV center and the cavity and the hyperfine interaction term  between a NV center and a singular nearest neighbour $13_C$
center. We propose the use of dressed state qubit \cite{11} and effective Hamiltonian theory \cite{18} for the cavity - $13_C$
Hamiltonian and perform simulation using QuTip \cite{22,23} and also do the analytical calculation to show the state transfer between the CPW cavity and $13_C$ center and Fidelity calculation
 to show how faithfully the state transfer can be performed. In the next section we elaborate on the theoretical model that was taken. Subsequently we show the results obtained by analytical calculation and verified by simulations.
 \section{Theoretical Description}
\begin{figure}[h]
    \centering
    \includegraphics[width=0.5\textwidth,height=5cm]{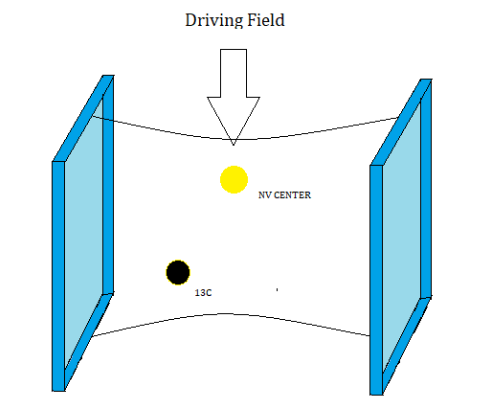}
    \caption{NV-13C system in cavity}
\end{figure}
In the following we consider a single $13_C$ center in close proximity to an NV center
in diamond inside a CPW cavity. To consider the coupling between these 3 systems we write the full Hamiltonian consisting of all 3 systems' zeroth order Hamiltonian and the interaction terms. The NV center's spin manifold has three possible spin states, $\ket{m_S=0}$, $\ket{m_S=-1}$, $\ket{m_S=1}$. In zero external magnetic field the later two are degenerate in energy, while $\ket{m_S=0}$ is lower in energy than the superposition of the later two states, owing to the Zero-field splitting introduced due to spin-spin interaction in the NV center. By applying a suitable external magnetic field the degeneracy between them can be broken. Due to this energy difference we can effectively encode our qubit state  using $\ket{m_S=0}$ and  $\ket{m_S=-1}$ states. Having coupled to a CPW cavity and driven by a Microwave radiation  tuned to the transition between $\ket{m_S=0}$ and  $\ket{m_S=-1}$ states, the Hamiltonian for the combined system reads \cite{11,19},
\begin{eqnarray}
H_{NV,c} = \frac{\omega_{NV}}{2} \sigma_z + \omega_c a^{\dagger}a + g(\ket{-1}\bra{0}a + \ket{0}\bra{-1} a^{\dagger}) + \lambda (\ket{-1}\bra{0}e^{-i\omega_0t} +H.C)\end{eqnarray}
where  $\sigma_z$ = $\ket{-1}\bra{-1} - \ket{0}\bra{0}$, $\omega_{NV}$ is the energy gap between $\ket{m_S=0}$ and $\ket{m_S=-1}$, given by $D_{ZF}$ - $\gamma_e$B, where B is applied magnetic field component that is parallel to NV spin quantization axis \cite{19}. $\omega_c$ is the cavity frequency, a($a^{\dagger}$) is the annihilation(creation) operator of the cavity modes, g the coupling frequency between NV and cavity, $\lambda$ the amplitude of the driving MW and $\omega_0$, the frequency of the driving field. We used the dressed state formalism as suggested in \cite{20} to simplify the Hamiltonian. \\
Now considering the NV center and $13_C $ interacion,
the Hamiltonian, for the NV-C interaction is given by \cite{21}
\begin{eqnarray}
\begin{aligned}
H_{NV,13_C} = \omega_{13_C} I_z + C_{\parallel}(\theta) \Sigma_z I_z + \\
& \frac{C_{\perp}(\theta)}{2} (\Sigma^{+} I_{-} +
\Sigma^{-} I_{+}) + \frac{C_{R}(\theta)}{2} (\Sigma^{+} I_{+}+
\Sigma^{-} I_{-}) + C_{\Delta}(\theta) (\Sigma_z I_y + \Sigma_y I_z)
\end{aligned}
\end{eqnarray}
Here $\omega_{13_C}$ = $\gamma_{13_C}$B (where B is magnetic field), $\Sigma_z$, $\Sigma^{+}$ and $\Sigma^{-}$ are the Pauli operators for the triplet spins and $I_z$, $I_{+}$, $I_{-}$ are the Pauli operators for the spin - 1/2 $13_C$ nuclei. Note here that $\Sigma^{+
}$ = ($\ket{+1}\bra{0} + \ket{0}\bra{-1} $). Out of the Coefficients we can ignore $C_R$ and $C_{\Delta}$ for high Magnetic field \cite{21}. The values for the rest of them are \cite{21}
\begin{eqnarray}
C_{\parallel}(\theta) = C_{\parallel} cos^{2}(\theta) + C_{\perp} sin^{2}(\theta) 
\\
C_{\perp}(\theta) = \frac{1}{2}(C_{\perp} (1+cos^2(\theta)) + C_{\parallel} sin^2(\theta))
\end{eqnarray}
where, $\theta$ is the angle between NV axis and vacancy-carbon axis. For the case we are considering we have removed the possibility of transition between the states $\ket{0}$ and $\ket{1}$. Hence our interaction Hamiltonian becomes,
\begin{eqnarray}
H_{NV,13_C} = \omega_{C13} I_z + C_{\parallel}(\theta) \ket{-1}\bra{-1} I_z + \frac{C_{\perp}(\theta)}{2} (\ket{0}\bra{-1} I_{-} + \ket{-1}\bra{0} I_{+})
\end{eqnarray}
Under the transformation (2) Equation (5) transforms as 
\begin{eqnarray}
\begin{aligned}
H_{NV,13_C}(e) = \omega_{C13} I_z + C_{\parallel}(\theta) \ket{-1}\bra{-1} I_z +\\
& \frac{C_{\perp}(\theta)}{2} (e^{-i\omega_0t}\ket{0}\bra{-1} I_{-} + e^{i\omega_0t}\ket{-1}\bra{0} I_{+})
\end{aligned}
\end{eqnarray}
and under the dressed state basis this can be written as
\begin{eqnarray}
\begin{aligned}
H^{'}_{NV,13_C}(e) = {} & \omega_{C13} I_z + \frac{C_{\parallel}(\theta)}{2} (I_z + \frac{cos(\eta)}{2}S_z I_z -\frac{sin(\eta)}{2} (S_{-}I_z+S_{+}I_z)) +\\
& \frac{C_{\perp}(\theta)}{2} (e^{-i\omega_0t}(\frac{sin(\eta)}{2} S_z + cos^2(\frac{eta}{2})S_{-}-sin^2(\frac{eta}{2})S_{+}) I_{-} + \\ 
& e^{i\omega_0t}(\frac{sin(\eta)}{2} S_z - sin^2(\frac{\eta}{2})S_{-}+cos^2(\frac{\eta}{2})S_{+}) I_{+})
\end{aligned}
\end{eqnarray}
Thus our dressed state Hamiltonian for the whole system is 
\begin{eqnarray}
\begin{aligned}
H_{tot} = {} &\frac{1}{2} \Omega S_z + \omega_c a^{\dagger}a + \frac{\kappa}{2} (S_{-}a^{\dagger} e^{-i\omega_0t} + h.c) + (\omega_{C13}+\frac{C_{\parallel}(\theta)}{2}) I_z + \\
& \frac{C_{\parallel}(\theta)}{2} ( \frac{cos(\eta)}{2}S_z I_z -\frac{sin(\eta)}{2} (S_{-}I_z+S_{+}I_z)) + \frac{C_{\perp}(\theta)}{2} (e^{-i\omega_0t}(\frac{sin(\eta)}{2} S_z + cos^2(\frac{\eta}{2})S_{-}-sin^2(\frac{\eta}{2})S_{+}) I_{-} \\
& + e^{i\omega_0t}(\frac{sin(\eta)}{2} S_z - sin^2(\frac{\eta}{2})S_{-}+cos^2(\frac{\eta}{2})S_{+}) I_{+})
\end{aligned}
\end{eqnarray}
Under the rotating wave approximation where we ignore the fast rotating terms, this gives us
\begin{eqnarray}
\begin{aligned}
H_I = {} & \frac{\kappa}{2}(S_{-}a^{\dagger}e^{i(\omega_c - \omega_0 - \Omega)t} + h.c) + \frac{C_{\parallel}(\theta)}{2} \frac{cos(\eta)}{2}S_z I_z -\\
& \frac{C_{\perp}(\theta)}{2} sin^2(\frac{\eta}{2}) S_{+} I_{-} e^{i(-\omega_0 - \omega^{'}_{13C} + \Omega)t} - \frac{C_{\perp}(\theta)}{2} sin^2(\frac{\eta}{2}) S_{-}I_{+} e^{i(\omega_0 + \omega^{'}_{13C} - \Omega)t}
\end{aligned}
\end{eqnarray}
where $\omega^{'}_{13C}$ = $2(\omega_{C13}+\frac{C_{\parallel}(\theta)}{2})$. Taking 
$\Delta_1$ = $\omega_c - \omega_0 - \Omega$ and $\Delta_2$ = $\omega_0 + \omega^{'}_{13C} - \Omega$; Equation (9) thus becomes
\begin{eqnarray}
\begin{aligned}
H_I = {} & \frac{\kappa}{2}(S_{-}a^{\dagger}e^{i\Delta_1t} + h.c) + \frac{C_{\parallel}(\theta)}{2} \frac{cos(\eta)}{2}S_z I_z +\\
& H (S_{+} I_{-} e^{-i\Delta_2t} + S_{-}I_{+} e^{i\Delta_2t})
\end{aligned}
\end{eqnarray}
where, H = $-\frac{C_{\perp}(\theta)}{2} sin^2(\frac{\eta}{2})$. Using the effective Hamiltonian theory \cite{5} with the above interaction Hamiltonian we obtain the effective Hamiltonian
\begin{eqnarray}
H_{eff} = \frac{\kappa^2}{4\omega_1} (-S_z a^{\dagger}a - \ket{e}\bra{e}) - \frac{\kappa H}{2\omega_{12}} S_z a^{\dagger}I_{-} e^{i\Delta_{12}t} - \frac{\kappa H}{2\omega_{12}} S_z aI_{+} e^{-i\Delta_{12}t} + \frac{H^2}{\omega_2}(S_{+}S_{-}I_z - S_z I_{+}I_{-})
\end{eqnarray}
here $\omega_i$ = $\Delta_i$, $\omega_{12} = \frac{2\omega_1\omega_2}{\omega_1+\omega_2}$ , $\Delta_{12}$ = $\omega_1 - \omega_2$. If we consider that our system was initially in $\ket{g}$ then the effective Hamiltonian is 
\begin{eqnarray}
H_{eff} = \frac{\kappa^2}{4\omega_1} a^{\dagger}a + \frac{\kappa H}{2\omega_{12}} a^{\dagger}I_{-}e^{i\Delta_{12}t} + 
\frac{\kappa H}{2\omega_{12}} aI_{+}e^{-i\Delta_{12}t} + \frac{H^2}{\omega_2} I_{+}I_{-}
\end{eqnarray}
It is clear from the formulation that by changing the values of $\Omega$, cavity mode frequency $\omega_c$ and the values of microwave field strength and frequency and $13_C$ we can control the values of $\omega_1$ and $\omega_2$, $\Delta_{12}$ and $\omega_{12}$.
\section{Results}
For a state transfer, we start with the initial state
\begin{eqnarray}
\ket{\psi}= cos\theta \ket{0}_c \ket{0}_I + sin\theta \ket{1}_c \ket{0}_I
\end{eqnarray}
where the subscripts c and I correspond to the Hilbert Spaces Cavity and Carbon atom respectively. Our target is to transfer this state into the $13_C$ state. We have effective Hamiltonian 
\begin{eqnarray}
H_{eff} = \frac{\kappa^2}{4\omega_1} a^{\dagger}a + \frac{\kappa H}{2\omega_{12}} a^{\dagger}I_{-}e^{i\Delta_{12}t} + 
\frac{\kappa H}{2\omega_{12}} aI_{+}e^{-i\Delta_{12}t} + \frac{H^2}{\omega_2} I_{+}I_{-}
\end{eqnarray}
Under this Hamiltonian at a time t, the state evolution at a time t can be expressed as 
\begin{eqnarray}
\ket{\psi(t)} = M_0(t) \ket{0}_c \ket{0}_I + M_1(t) \ket{1}_c \ket{0}_I + M_2(t) \ket{0}_c \ket{1}_I
\end{eqnarray}
These coefficients can be calculated from the Schrodinger equation.
\begin{eqnarray} 
i\frac{d \ket{\psi(t)}}{dt} = H_{eff} \ket{\psi(t)}
\end{eqnarray}
\begin{eqnarray} 
i( \dot{M_0}\ket{0}_c \ket{0}_I + \dot{M_1} \ket{1}_c \ket{0}_I + \dot{M_2} \ket{0}_c \ket{1}_I)=H_e\ket{\psi(t)}
\end{eqnarray}
This leads to the coupled differential equations,
\begin{eqnarray}
\dot{M_0}=0
\\
i\dot{M_1} = \frac{\kappa^2}{4\omega_1} M_1 + \frac{\kappa H}{2\omega_{12}}e^{i\Delta_{12}t} M_2
\\
i\dot{M_2} = \frac{\kappa H}{2\omega_{12}}e^{-i\Delta_{12}t} M_1 + \frac{H^2}{\omega_2} M_2
\end{eqnarray}
The expressions for $M_1$,$M_2$,$M_0$ can be calculated by solving the secular equation which when written in determinant form reads,
\\
\\
$\begin{vmatrix}
\frac{-i \kappa^2}{4\omega_1} & \frac{-i \kappa H e^{i\Delta_{12}t}}{2 \omega_{12}} 
\\
\\
\frac{-i \kappa H e^{-i\Delta_{12}t}}{2 \omega_{12}} & \frac{-i  H^2}{\omega_2}
\end{vmatrix}$
=
0
\\
\\
To observe the state population evolution and inspect the state transfer, we simulated this Hamiltonian in QuTip with the given initial state and did the Fidelity calculation and subsequently we also solved the problem analytically. The two results match with each other remarkably. Here in the next subsection we first present the analytical solution. 

\subsection{Analytical Solution}
For analytical solution we solved the matrix given by (27) and (28) for their eigenvalues. Eigenvalues and their corresponding eigenvectors are given 
\begin{eqnarray}
\lambda_1 = i(-\frac{\kappa^2}{8 \omega_1} - \frac{H^2}{2\omega_2})-i\sqrt{\frac{\kappa^4}{64\omega_1^2}-\frac{\kappa^2H^2}{8\omega_1\omega_2}+\frac{\kappa^2H^2}{4\omega_{12}^2}+\frac{H^4}{4\omega_2^2}}
\\
\ket{\lambda_1} = \begin{bmatrix} \frac{\frac{\kappa He^{i\Delta_{12}t}}{\omega_{12}}}{(\frac{H^2}{\omega_2} - \frac{\kappa^2}{4\omega_1})+\sqrt{\frac{\kappa^4}{16\omega_1^2}-\frac{\kappa^2H^2}{2\omega_1\omega_2}+\frac{\kappa^2H^2}{\omega_{12}^2}+\frac{H^4}{\omega_2^2}}}
\\
\\
\end{bmatrix}
\\
\lambda_2 = i(-\frac{\kappa^2}{8 \omega_1} - \frac{H^2}{2\omega_2})+i\sqrt{\frac{\kappa^4}{64\omega_1^2}-\frac{\kappa^2H^2}{8\omega_1\omega_2}+\frac{\kappa^2H^2}{4\omega_{12}^2}+\frac{H^4}{4\omega_2^2}}
\\
\ket{\lambda_2} = \begin{bmatrix} \frac{\frac{\kappa He^{i\Delta_{12}t}}{\omega_{12}}}{(\frac{H^2}{\omega_2} - \frac{\kappa^2}{4\omega_1})-\sqrt{\frac{\kappa^4}{16\omega_1^2}-\frac{\kappa^2H^2}{2\omega_1\omega_2}+\frac{\kappa^2H^2}{\omega_{12}^2}+\frac{H^4}{\omega_2^2}}}
\\
\\

\end{bmatrix}
\end{eqnarray}
And subsequently the matrix 
$\begin{bmatrix} M_1 \\ M_2 \end{bmatrix}$can be found as
\begin{eqnarray}
\begin{bmatrix} M_1 \\ M_2 \end{bmatrix} = \alpha\ket{\lambda_1}e^{\lambda_1t} + \beta\ket{\lambda_2}e^{\lambda_2t}
\end{eqnarray}
With
\\
\begin{eqnarray}
M_1 = \alpha \frac{\frac{\kappa He^{i\Delta_{12}t}}{\omega_{12}}}{(\frac{H^2}{\omega_2} - \frac{\kappa^2}{4\omega_1})+\sqrt{\frac{\kappa^4}{16\omega_1^2}-\frac{\kappa^2H^2}{2\omega_1\omega_2}+\frac{\kappa^2H^2}{\omega_{12}^2}+\frac{H^4}{\omega_2^2}}} e^{\lambda_1t} + \beta \frac{\frac{\kappa He^{i\Delta_{12}t}}{\omega_{12}}}{(\frac{H^2}{\omega_2} - \frac{\kappa^2}{4\omega_1})-\sqrt{\frac{\kappa^4}{16\omega_1^2}-\frac{\kappa^2H^2}{2\omega_1\omega_2}+\frac{\kappa^2H^2}{\omega_{12}^2}+\frac{H^4}{\omega_2^2}}} e^{\lambda_2t}\end{eqnarray}
\\
\begin{eqnarray}
M_2 = \alpha e^{\lambda_1t} + \beta e^{\lambda_2t}
\end{eqnarray}
From initial conditions, we can determine $\alpha$ and $\beta$. Since at t=0, $M_2$ = 0, $\alpha$=-$\beta$
and $M_1$ = $sin\theta$, which gives 
\begin{eqnarray}
\beta = \frac{-\kappa H sin\theta}{4\omega_{12}D}
\\
D = \sqrt{\frac{\kappa^4}{64\omega_1^2}-\frac{\kappa^2H^2}{8\omega_1\omega_2}+\frac{\kappa^2H^2}{4\omega_{12}^2}+\frac{H^4}{4\omega_2^2}}
\end{eqnarray}
from which $M_1$ and $M_2$ are expressed as
\begin{eqnarray}
M_1(t) = \frac{1}{4D} (4D cos(Dt) + 2i(\frac{H^2}{\omega_2} - \frac{\kappa^2}{4\omega_1}) sin(Dt)) e^{i(C+\Delta_{12})t} sin\theta
\end{eqnarray}
\begin{eqnarray}
M_2(t) = -\frac{i\kappa He^{iCt}}{2\omega_{12}D} sin(Dt) sin\theta
\end{eqnarray}
\begin{eqnarray}
C = -\frac{\kappa^2}{8\omega_1} - \frac{H^2}{2\omega_2}
\end{eqnarray}
The $\omega$s here are detunings, g, H are coupling coefficients. Now according to the coupling coefficients, we can change the detunings to our choice (and even one $\omega$ according to the other) so that $\frac{H^2}{\omega_2}$ = $\frac{\kappa^2}{4\omega_1}$ . This way the imaginary part of $M_1$ vanish. Notice that the final state we obtain is \begin{eqnarray}
\ket{\psi(\tau)}= cos\theta \ket{0}_c \ket{0}_I + i \hspace{0.1cm}sin\theta \ket{0}_c \ket{1}_I
\end{eqnarray} from which the initial state can be obtained using a unitary transformation.\\
The values of $\kappa$ and H can be determined experimentally. For our choice I took $\kappa$ = 1000MHz, H = -32.02 MHz $\omega_1$=2MHz, $\omega_2$ = 0.008 MHz. We took $\theta$ = $\frac{\pi}{6}$. Later changing the values of $\theta$ the following graphs(Figs 2-5) were obtained for population evolution of time.\\
\begin{figure}[htb]

\begin{minipage}{0.6\textwidth}
\begin{tikzpicture}
  \node (img)  {\includegraphics[scale=0.65]{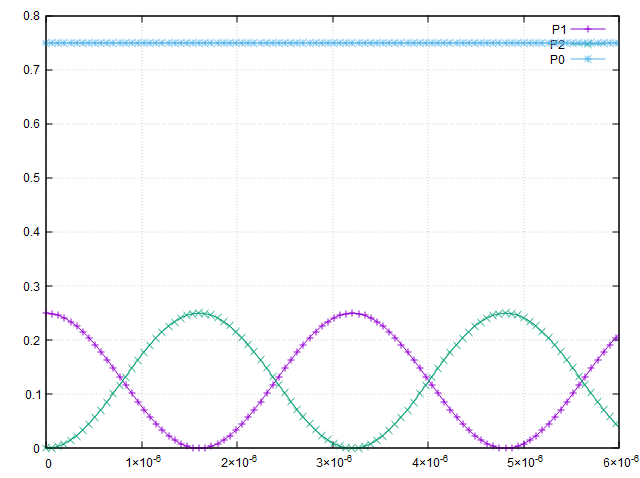}};
  \node[below=of img, node distance=0cm, yshift=1cm,font=\color{black}] {time /sec};
  \node[left=of img, node distance=0cm, rotate=90, anchor=center,yshift=-0.7cm,font=\color{black}] {Population};
 \end{tikzpicture}
\end{minipage}%
\caption{Time evolution plot for $\theta$=$\frac{\pi}{6}$}
\end{figure}
\begin{figure}
\begin{minipage}{0.6\textwidth}
\begin{tikzpicture}
  \node (img)  {\includegraphics[scale=0.65]{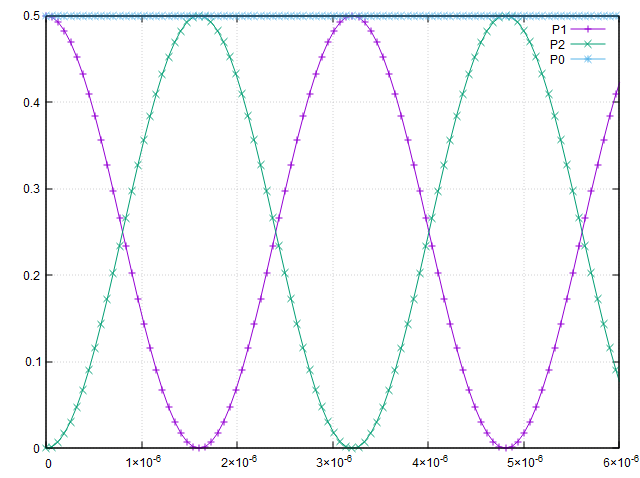}};
  \node[below=of img, node distance=0cm, yshift=1cm,font=\color{black}] {time/sec};
  \node[left=of img, node distance=0cm, rotate=90, anchor=center,yshift=-0.7cm,font=\color{black}] {Population};

 \end{tikzpicture}

\end{minipage}%
\caption{Time evolution plot for $\theta$=$\frac{\pi}{4}$ }
\end{figure}
\begin{figure}[htb]
\begin{minipage}{0.6\textwidth}
\begin{tikzpicture}
  \node (img)  {\includegraphics[scale=0.65]{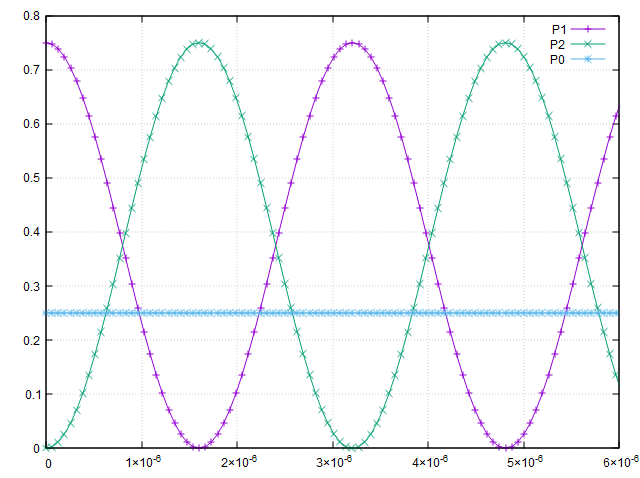}};
  \node[below=of img, node distance=0cm, yshift=1cm,font=\color{black}] {time/sec};
  \node[left=of img, node distance=0cm, rotate=90, anchor=center,yshift=-0.7cm,font=\color{black}] {Population};
\end{tikzpicture}
\end{minipage}%
\caption{Time evolution plot for $\theta$=$\frac{\pi}{3}$ }
\end{figure}
\begin{figure}
\begin{minipage}{0.6\textwidth}
\begin{tikzpicture}
  \node (img)  {\includegraphics[scale=0.65]{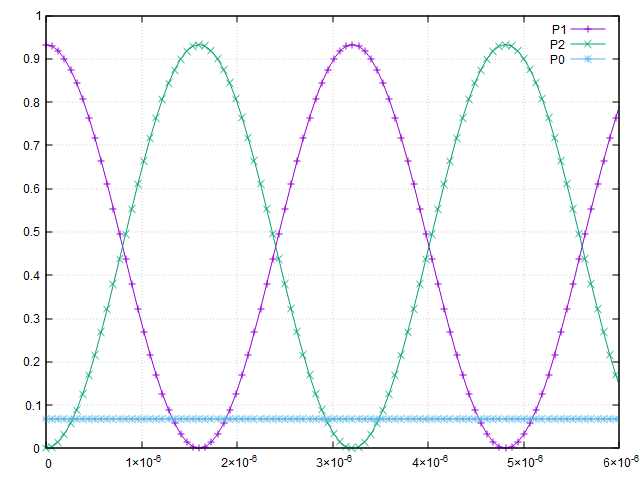}};
  \node[below=of img, node distance=0cm, yshift=1cm,font=\color{black}] {time/sec};
  \node[left=of img, node distance=0cm, rotate=90, anchor=center,yshift=-0.7cm,font=\color{black}] {Population};
\end{tikzpicture}
\end{minipage}%
\caption{Time evolution of population plots for  $\theta$=75 }
\end{figure}

    
    

    

\pagebreak
Here the terms P0,P1,P2 notifies the population in $\ket{0}_c \ket{0}_I$,$\ket{1}_c \ket{0}_I$ and $\ket{0}_c \ket{1}_I$ respectively. Clearly P0 remains constant noting that the first state's population remains the same. As can be seen from the graphs as well as calculations that in 1.603x$10^{-6}$ seconds there is a clear population transfer from $\ket{1}_c \ket{0}_I$ to $\ket{0}_c \ket{1}_I$, where P1 goes to 0. Also note that this value of time does not depend on the angles $\theta$, but rather depends on the coupling coefficients and the detunings, signifying that this time is same for all values of $\theta$ for the state that is to be transferred.
\\
\clearpage
\subsubsection{Fidelity Calculation}
Dephasing in the cavity and $13_C$  atoms can be introduced through the dephasing time and consequently the dephasing constant in the effective Hamiltonian, leading to the final hamiltonian as suggested in \cite{11}
\begin{eqnarray}
H_e = \frac{\kappa^2}{4\omega_1} a^{\dagger} a + \frac{\kappa H}{2\omega_{12}}a^{\dagger} I^{-}e^{i\Delta_{12}t}+ \frac{\kappa H}{2\omega_{12}} a I^{+}e^{-i\Delta_{12}t} + \frac{H^2}{\omega_2} I^{+}I^{-} - i \frac{k_1}{2} a^{\dagger}a - i\frac{k_2}{2} I^{+}I^{-}
\end{eqnarray}
where $k_1$ and $k_2$ are dephasing constants for cavity and carbon dephasing. Conditional Fidelity calculation can then be done using the reduced density matrix of the Carbon atom by tracing out the cavity part. It is given by
\begin{eqnarray}
\rho_{I} = Tr_c(\ket{\psi(t)}\bra{\psi(t)})
\end{eqnarray}
At a time when state transfer occurs, the final state of 13C would be $\psi_{\tau}$=$cos\theta \ket{0} + i\hspace{0.1cm} sin\theta \ket{1}$. This means, the conditional fidelity as a function of time is 
\begin{eqnarray}
F = \bra{\psi_{\tau}}\rho_I\ket{\psi_{\tau}}
\end{eqnarray},
For the Hamiltonian with dephasing, the expressions for $M_0$,$M_1$,$M_2$ were calculated as before,
except that this time the dephasing constants were included in the coupled differential equation. Same values of the parameters as the previous instance were chosen for this case. The expressions for these are---
\begin{eqnarray}
M'_1(t) = e^{At}e^{i\Delta_{12}t}(\frac{C}{B'} sin(\frac{B't}{2}) + cos(\frac{B't}{2})) sin\theta
\\
M'_2(t) = \frac{-i\kappa H}{B'\omega_{12}} e^{At} sin(\frac{B't}{2}) sin\theta
\\
A = -i\frac{\kappa^2}{8\omega_1} - \frac{\kappa_1}{2} - -i\frac{H^2}{2\omega_2} - \frac{\kappa_2}{2}
\\
B' = \sqrt{\frac{\kappa^2H^2}{\omega_{12}} - \frac{(\kappa_1 - \kappa_2)^2}{4}}
\\
C = \frac{\kappa_2 - \kappa_1}{2}
\end{eqnarray}
 Fidelities were calculated for all $\theta$s. The plots for fidelity of the state transfer process as a function of time for different $\theta$s are given below
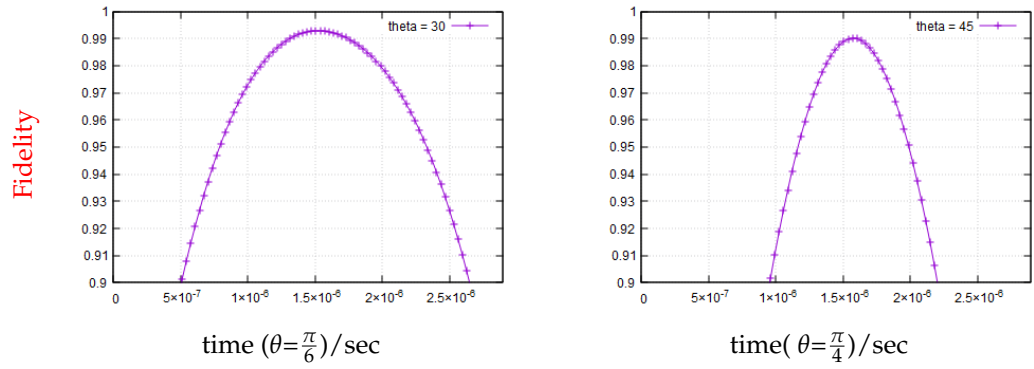
\begin{figure}[htb]

\begin{minipage}{0.5\textwidth}
\begin{tikzpicture}
  \node (img)  {\includegraphics[scale=0.53]{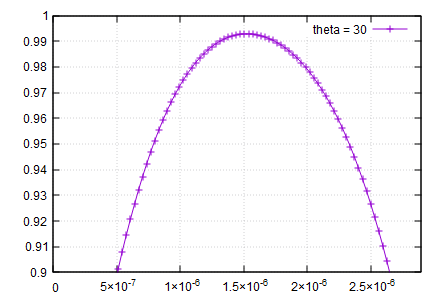}};
  \node[below=of img, node distance=0cm, yshift=1cm,font=\color{black}] {time ($\theta$=$\frac{\pi}{6}$)/sec};
  \node[left=of img, node distance=0cm, rotate=90, anchor=center,yshift=-0.7cm,font=\color{red}] {Fidelity};
 \end{tikzpicture}
\end{minipage}
\begin{minipage}{0.5\textwidth}
\begin{tikzpicture}
  \node (img)  {\includegraphics[scale=0.53]{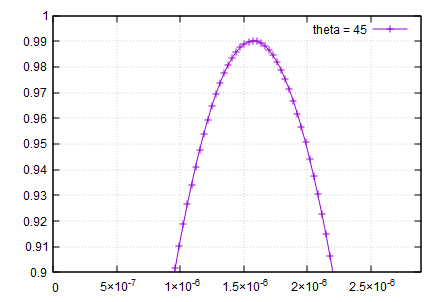}};
  \node[below=of img, node distance=0cm, yshift=1cm,font=\color{black}] {time( $\theta$=$\frac{\pi}{4}$)/sec};

 \end{tikzpicture}

\end{minipage}%
\caption{Fidelity plots for $\theta$=$\frac{\pi}{4}$ and $\theta$=$\frac{\pi}{6}$ as a function of time}
\end{figure}

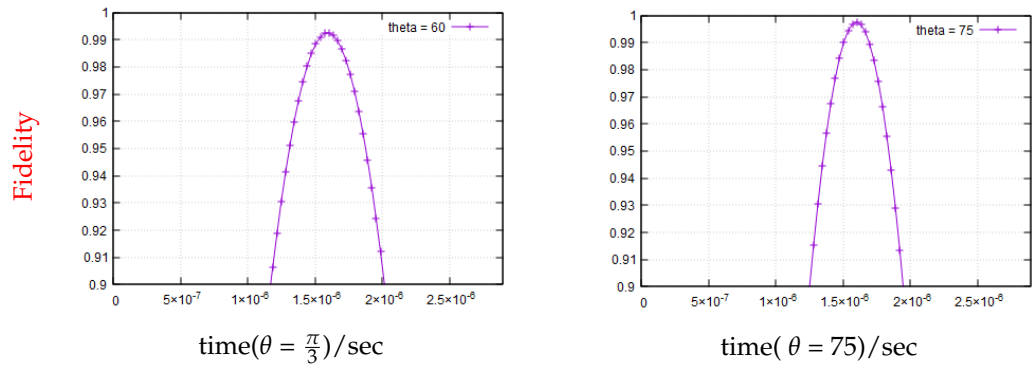
\begin{figure}[htb]
\begin{minipage}{0.5\textwidth}
\begin{tikzpicture}
  \node (img)  {\includegraphics[scale=0.53]{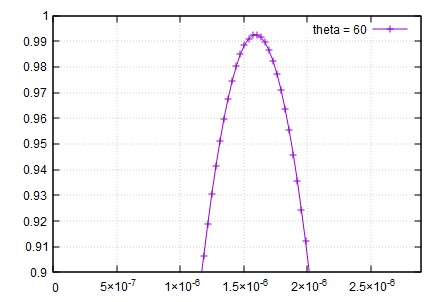}};
  \node[below=of img, node distance=0cm, yshift=1cm,font=\color{black}] {time($\theta$ = $\frac{\pi}{3}$)/sec};
  \node[left=of img, node distance=0cm, rotate=90, anchor=center,yshift=-0.7cm,font=\color{red}] {Fidelity};
 \end{tikzpicture}
\end{minipage}%
\begin{minipage}{0.5\textwidth}
\begin{tikzpicture}
  \node (img)  {\includegraphics[scale=0.53]{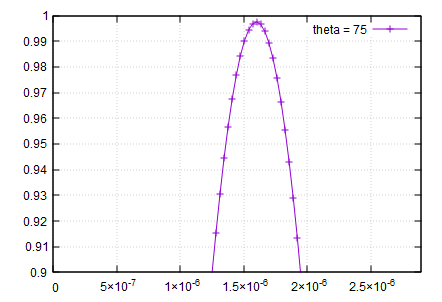}};
  \node[below=of img, node distance=0cm, yshift=1cm,font=\color{black}] {time( $\theta$ = 75)/sec};
\end{tikzpicture}
\end{minipage}%
\caption{Fidelity plots for $\theta$=$\frac{\pi}{3}$ and $\theta$=75 }
\end{figure}

    

\begin{figure}
\begin{minipage}{0.5\textwidth}
\begin{tikzpicture}
    \centering
    \node (img)  {\includegraphics[scale=0.64]{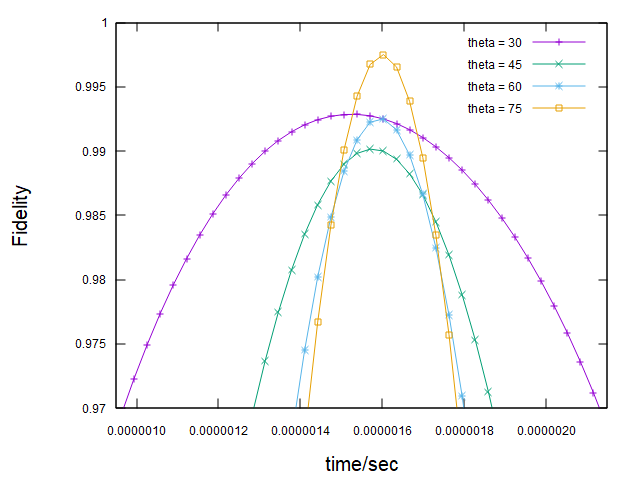}};
\end{tikzpicture} 
\end{minipage}%
\caption{Fidelity plots for all thetas together}
\end{figure}
They are all shown to reach the minimum fidelity of 0.990 within the operation time of 1.602 microseconds for the state transfer as can be seen from Fig 8. The fidelity is least for $\theta$ = $\frac{\pi}{4}$.
\clearpage
\subsection{QuTip Simulations}
We used QuTip \cite{22,23} to simulate the effective Hamiltonian given by eqn (21) as the initial state with the qutip.mesolver  function. We solved this for the same $\theta$s and the resultant population evolution plots(Fig.9-12) as functions of time are given below.
\begin{figure}[h]
\centering
\begin{minipage}{0.5\textwidth}
\begin{tikzpicture}
  \node (img)  {\includegraphics[scale=0.6]{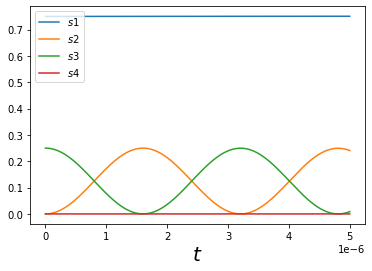}};
  \node[below=of img, node distance=0cm, yshift=1cm,font=\color{black}] {time /sec};
  \node[left=of img, node distance=0cm, rotate=90, anchor=center,yshift=-0.7cm,font=\color{black}] {Population};
 \end{tikzpicture}
\end{minipage}%
\caption{Time evolution plot for $\theta$=$\frac{\pi}{6}$}
\end{figure}
\begin{figure}[h]
\centering
\begin{minipage}{0.5\textwidth}
\begin{tikzpicture}
  \node (img)  {\includegraphics[scale=0.6]{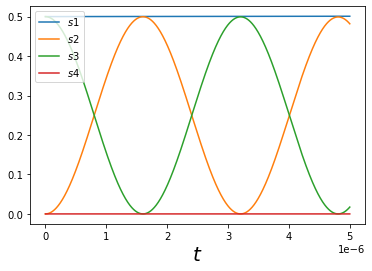}};
  \node[below=of img, node distance=0cm, yshift=1cm,font=\color{black}] {time/sec};
  \node[left=of img, node distance=0cm, rotate=90, anchor=center,yshift=-0.7cm,font=\color{black}] {Population};

 \end{tikzpicture}

\end{minipage}%
\caption{Time evolution plot for $\theta$=$\frac{\pi}{4}$ }
\end{figure}
\begin{figure}[h]
\centering
\begin{minipage}{0.5\textwidth}
\begin{tikzpicture}
  \node (img)  {\includegraphics[scale=0.6]{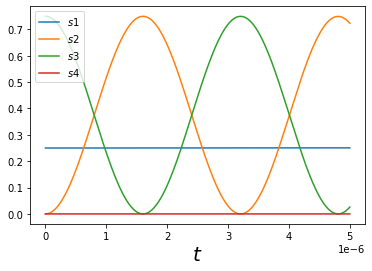}};
  \node[below=of img, node distance=0cm, yshift=1cm,font=\color{black}] {time/sec};
  \node[left=of img, node distance=0cm, rotate=90, anchor=center,yshift=-0.7cm,font=\color{black}] {Population};
\end{tikzpicture}
\end{minipage}%
\caption{Time evolution plot for $\theta$=$\frac{\pi}{3}$ }
\end{figure}
\begin{figure}[h]
\centering
\begin{minipage}{0.5\textwidth}
\begin{tikzpicture}
  \node (img)  {\includegraphics[scale=0.6]{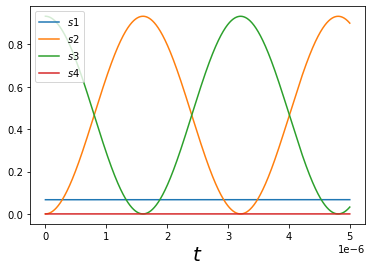}};
  \node[below=of img, node distance=0cm, yshift=1cm,font=\color{black}] {time/sec};
  \node[left=of img, node distance=0cm, rotate=90, anchor=center,yshift=-0.7cm,font=\color{black}] {Population};
\end{tikzpicture}
\end{minipage}%
\caption{Time evolution of population plots for  $\theta$=75 }
\end{figure}
  
       
    
    
    
    
    

\clearpage
In all cases, s1 denotes the probability of being in state $\ket{0}_c\ket{0}_I$, s2 denotes that for the state $\ket{0}_c\ket{1}_I$, s3 for $\ket{1}_c\ket{0}_I$, s4 for $\ket{1}_c\ket{1}_I$. s4 is identically zero in all cases, as expected. At an elapsed time of 1.603x$10^{-6}$ seconds there is swap of population between s1 and s2. The fidelities for this with the decay terms were found using the master equation solver with the collapse operators for cavity and $13_C$ by solving the Lindblad Master Equation. The fidelities of the state transfer as a function of time for different $\theta$s were plotted and the graphs are given in(Fig. 13-14)
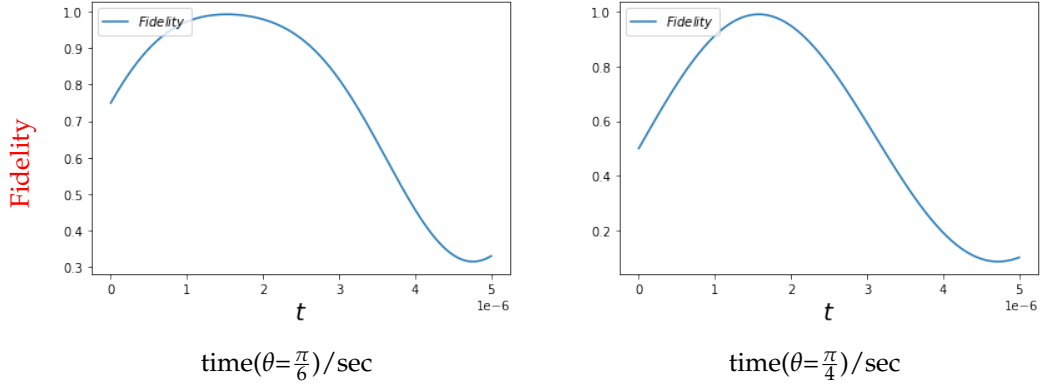
\begin{figure}[htb]
\begin{minipage}{0.5\textwidth}
\begin{tikzpicture}
  \node (img)  {\includegraphics[scale=0.47]{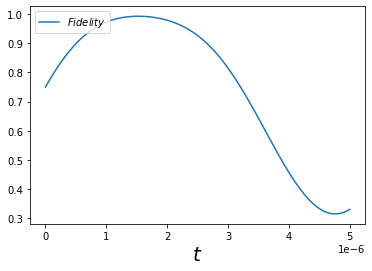}};
  \node[below=of img, node distance=0cm, yshift=1cm,font=\color{black}] {time($\theta$=$\frac{\pi}{6}$)/sec};
  \node[left=of img, node distance=0cm, rotate=90, anchor=center,yshift=-0.7cm,font=\color{red}] {Fidelity};
 \end{tikzpicture}
\end{minipage}%
\begin{minipage}{0.5\textwidth}
\begin{tikzpicture}
  \node (img)  {\includegraphics[scale=0.47]{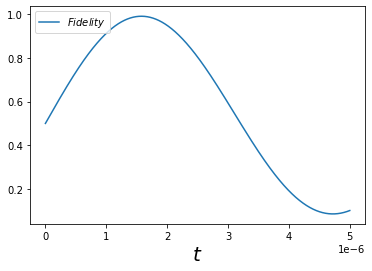}};
  \node[below=of img, node distance=0cm, yshift=1cm,font=\color{black}] {time($\theta$=$\frac{\pi}{4}$)/sec};

\end{tikzpicture}
\end{minipage}%
\caption{Fidelity as a function of time for $\theta$ = $\frac{\pi}{6}$ and $\frac{\pi}{4}$}
\end{figure}
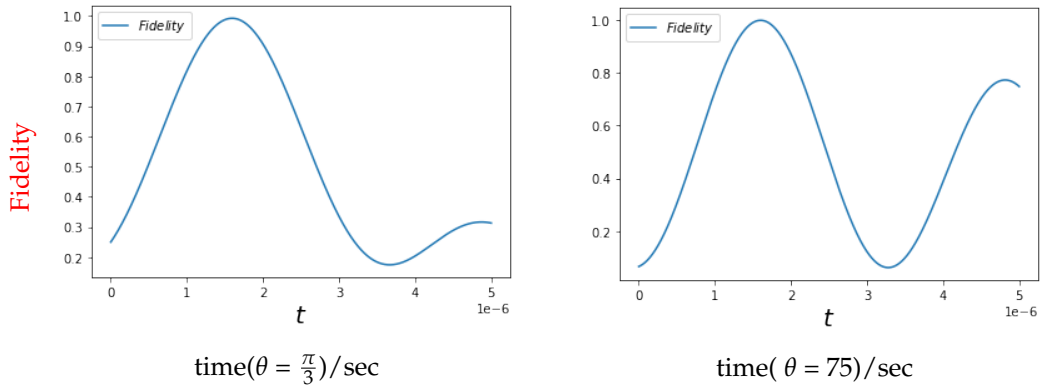
\begin{figure}[htb]
\begin{minipage}{0.5\textwidth}
\begin{tikzpicture}
  \node (img)  {\includegraphics[scale=0.47]{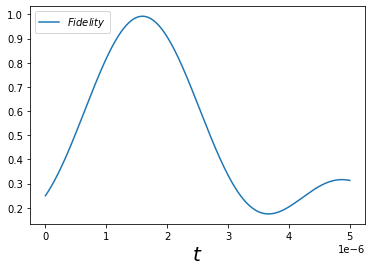}};
  \node[below=of img, node distance=0cm, yshift=1cm,font=\color{black}] {time($\theta$ = $\frac{\pi}{3}$)/sec};
  \node[left=of img, node distance=0cm, rotate=90, anchor=center,yshift=-0.7cm,font=\color{red}] {Fidelity};
 \end{tikzpicture}
\end{minipage}%
\begin{minipage}{0.5\textwidth}
\begin{tikzpicture}
  \node (img)  {\includegraphics[scale=0.47]{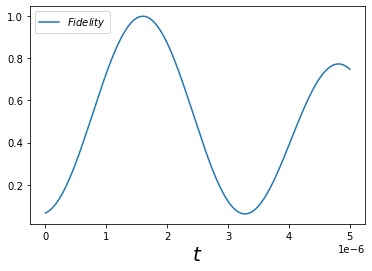}};
  \node[below=of img, node distance=0cm, yshift=1cm,font=\color{black}] {time( $\theta$ = 75)/sec};
\end{tikzpicture}
\end{minipage}%
\caption{Fidelity as a function of time for $\theta$ = $\frac{\pi}{3}$ and 75}
\end{figure}

    
    

At an elapsed time of 1.603x$10^{-6}$seconds (approximately) the fidelity of state transfer for $\theta = \frac{\pi}{6}$ is 0.992, for $\theta = \frac{\pi}{4}$ is 0.990, for $\theta = \frac{\pi}{3}$ is 0.992 and for $\theta = \frac{75\pi}{180}$ is 0.997.

\section{Conclusion}
In conclusion we have provided a theoretical framework for a faithful quantum state transfer protocol between a CPW cavity and $13_C$ nuclear spin. We have calculated the interaction hamiltonian for cavity-NV center-$13_C$ system and used the effective hamiltonian theory to formulate an effective hamiltonian and simulated that using QuTip as well as analytical methods. This gives us a fidelity of over 0.990 for the state transfer. From the analytical calculation it is clear that by tuning the detunings $\omega_1$ and $\omega_2$ and thereby tuning the Rabi frequency and the cavity frequency in correspondence with the coupling constants a direct state transfer is possible from CPW cavity to $13_C$ nucleus with good fidelity. This essentially gives us a detailed description of a quantum information transfer between the flying qubit of photon-cavity and the stationary qubit of $13_C$ which is a necessary process in a node in a quantum network. Also through further extension of the hyperfine interaction between a second NV center - $13_C$ system it has the potential to be used in a multi-qubit quantum memory.

\medskip
\printbibliography

@misc{1,
    title={Quantum State Transfer with Spin Chains},
    author={Daniel Burgarth},
    year={2007},
    eprint={0704.1309},
    archivePrefix={arXiv},
    primaryClass={quant-ph}
}

@article{2,
   title={Quantum State Transfer and Entanglement Distribution among Distant Nodes in a Quantum Network},
   volume={78},
   ISSN={1079-7114},
   url={http://dx.doi.org/10.1103/PhysRevLett.78.3221},
   DOI={10.1103/physrevlett.78.3221},
   number={16},
   journal={Physical Review Letters},
   publisher={American Physical Society (APS)},
   author={Cirac, J. I. and Zoller, P. and Kimble, H. J. and Mabuchi, H.},
   year={1997},
   month={Apr},
   pages={3221–3224}
}

@article{3,
	doi = {10.1088/1674-1056/27/2/024203},
	url = {https://doi.org/10.1088/1674-1056/27/2/024203},
	year = 2018,
	month = {feb},
	publisher = {{IOP} Publishing},
	volume = {27},
	number = {2},
	pages = {024203},
	author = {Pei Pei and He-Fei Huang and Yan-Qing Guo and Xing-Yuan Zhang and Jia-Feng Dai},
	title = {Quantum state transfer via a hybrid solid{\textendash}optomechanical interface},
	journal = {Chinese Physics B}
}

@article {4,
	Title = {Coherent quantum state storage and transfer between two phase qubits via a resonant cavity},
	Author = {Sillanpää, Mika A and Park, Jae I and Simmonds, Raymond W},
	DOI = {10.1038/nature06124},
	Number = {7161},
	Volume = {449},
	Month = {September},
	Year = {2007},
	Journal = {Nature},
	ISSN = {0028-0836},
	Pages = {438—442},
	URL = {https://doi.org/10.1038/nature06124},
}

@article{5,
	doi = {10.1088/1367-2630/14/9/093040},
	url = {https://doi.org/10.1088/1367-2630/14/9/093040},
	year = 2012,
	month = {sep},
	publisher = {{IOP} Publishing},
	volume = {14},
	number = {9},
	pages = {093040},
	author = {A Ruschhaupt and Xi Chen and D Alonso and J G Muga},
	title = {Optimally robust shortcuts to population inversion in two-level quantum systems},
	journal = {New Journal of Physics}
}

@article{6,
   title={Shortcuts to adiabatic passage for fast generation of Greenberger-Horne-Zeilinger states by transitionless quantum driving},
   volume={5},
   ISSN={2045-2322},
   url={http://dx.doi.org/10.1038/srep15616},
   DOI={10.1038/srep15616},
   number={1},
   journal={Scientific Reports},
   publisher={Springer Science and Business Media LLC},
   author={Chen, Ye-Hong and Xia, Yan and Song, Jie and Chen, Qing-Qin},
   year={2015},
   month={Oct}
}

@article{7,
  title = {Generating nonclassical photon states via longitudinal couplings between superconducting qubits and microwave fields},
  author = {Zhao, Yan-Jun and Liu, Yu-Long and Liu, Yu-xi and Nori, Franco},
  journal = {Phys. Rev. A},
  volume = {91},
  issue = {5},
  pages = {053820},
  numpages = {15},
  year = {2015},
  month = {May},
  publisher = {American Physical Society},
  doi = {10.1103/PhysRevA.91.053820},
  url = {https://link.aps.org/doi/10.1103/PhysRevA.91.053820}
}

@article{8,
   title={Hamiltonian engineering for robust quantum state transfer and qubit readout in cavity QED},
   volume={19},
   ISSN={1367-2630},
   url={http://dx.doi.org/10.1088/1367-2630/aa5d33},
   DOI={10.1088/1367-2630/aa5d33},
   number={2},
   journal={New Journal of Physics},
   publisher={IOP Publishing},
   author={Beaudoin, Félix and Blais, Alexandre and Coish, W A},
   year={2017},
   month={Feb},
   pages={023041}
}

@article{9,
   title={Quantum-state transfer from an ion to a photon},
   volume={7},
   ISSN={1749-4893},
   url={http://dx.doi.org/10.1038/nphoton.2012.358},
   DOI={10.1038/nphoton.2012.358},
   number={3},
   journal={Nature Photonics},
   publisher={Springer Science and Business Media LLC},
   author={Stute, A. and Casabone, B. and Brandstätter, B. and Friebe, K. and Northup, T. E. and Blatt, R.},
   year={2013},
   month={Feb},
   pages={219–222}
}

@article{10,
  title = {Quantum dynamics and quantum state transfer between separated nitrogen-vacancy centers embedded in photonic crystal cavities},
  author = {Yang, W. L. and Yin, Z. Q. and Xu, Z. Y. and Feng, M. and Oh, C. H.},
  journal = {Phys. Rev. A},
  volume = {84},
  issue = {4},
  pages = {043849},
  numpages = {10},
  year = {2011},
  month = {Oct},
  publisher = {American Physical Society},
  doi = {10.1103/PhysRevA.84.043849},
  url = {https://link.aps.org/doi/10.1103/PhysRevA.84.043849}
}

@article{11,
title = {High-fidelity quantum state transfer and strong coupling in a hybrid NV center coupled to CPW cavity system},
journal = {Chinese Journal of Physics},
volume = {66},
pages = {9-14},
year = {2020},
issn = {0577-9073},
doi = {https://doi.org/10.1016/j.cjph.2020.02.035},
url = {https://www.sciencedirect.com/science/article/pii/S0577907320301155},
author = {Qinghong Liao and Yanchao Fu and Jiangong Hu}
}

@article{12,
author = {Kennedy,T. A.  and Colton,J. S.  and Butler,J. E.  and Linares,R. C.  and Doering,P. J. },
title = {Long coherence times at 300 K for nitrogen-vacancy center spins in diamond grown by chemical vapor deposition},
journal = {Applied Physics Letters},
volume = {83},
number = {20},
pages = {4190-4192},
year = {2003},
doi = {10.1063/1.1626791},

URL = { 
        https://doi.org/10.1063/1.1626791
    
},
eprint = { 
        https://doi.org/10.1063/1.1626791
    
}

}

@PHDTHESIS{13,
       author = {{Childress}, Lilian Isabel},
        title = "{Coherent manipulation of single quantum systems in the solid state}",
       school = {Harvard University},
         year = 2007,
        month = dec,
       adsurl = {https://ui.adsabs.harvard.edu/abs/2007PhDT........99C}
}

@article{14,
  title = {Fault-Tolerant Quantum Communication Based on Solid-State Photon Emitters},
  author = {Childress, L. and Taylor, J. M. and S\o{}rensen, A. S. and Lukin, M. D.},
  journal = {Phys. Rev. Lett.},
  volume = {96},
  issue = {7},
  pages = {070504},
  numpages = {4},
  year = {2006},
  month = {Feb},
  publisher = {American Physical Society},
  doi = {10.1103/PhysRevLett.96.070504},
  url = {https://link.aps.org/doi/10.1103/PhysRevLett.96.070504}
}

@article{15,
   title={Fault-tolerant quantum repeaters with minimal physical resources and implementations based on single-photon emitters},
   volume={72},
   ISSN={1094-1622},
   url={http://dx.doi.org/10.1103/PhysRevA.72.052330},
   DOI={10.1103/physreva.72.052330},
   number={5},
   journal={Physical Review A},
   publisher={American Physical Society (APS)},
   author={Childress, L. and Taylor, J. M. and Sørensen, A. S. and Lukin, M. D.},
   year={2005},
   month={Nov}
}

@article {16,
	author = {Dutt, M. V. Gurudev and Childress, L. and Jiang, L. and Togan, E. and Maze, J. and Jelezko, F. and Zibrov, A. S. and Hemmer, P. R. and Lukin, M. D.},
	title = {Quantum Register Based on Individual Electronic and Nuclear Spin Qubits in Diamond},
	volume = {316},
	number = {5829},
	pages = {1312--1316},
	year = {2007},
	doi = {10.1126/science.1139831},
	publisher = {American Association for the Advancement of Science},
	issn = {0036-8075},
	URL = {https://science.sciencemag.org/content/316/5829/1312},
	eprint = {https://science.sciencemag.org/content/316/5829/1312.full.pdf},
	journal = {Science}
}

@article{17,
   title={Cavity QED with Magnetically Coupled Collective Spin States},
   volume={107},
   ISSN={1079-7114},
   url={http://dx.doi.org/10.1103/PhysRevLett.107.060502},
   DOI={10.1103/physrevlett.107.060502},
   number={6},
   journal={Physical Review Letters},
   publisher={American Physical Society (APS)},
   author={Amsüss, R. and Koller, Ch. and Nöbauer, T. and Putz, S. and Rotter, S. and Sandner, K. and Schneider, S. and Schramböck, M. and Steinhauser, G. and Ritsch, H. and et al.},
   year={2011},
   month={Aug}
}

@article{18,
   title={Effective Hamiltonian theory and its applications in quantum information},
   volume={85},
   ISSN={1208-6045},
   url={http://dx.doi.org/10.1139/P07-060},
   DOI={10.1139/p07-060},
   number={6},
   journal={Canadian Journal of Physics},
   publisher={Canadian Science Publishing},
   author={James, D F and Jerke, J},
   year={2007},
   month={Jun},
   pages={625–632}
}

@article{19,
       author = {{Hua}, Ming and {Tao}, Ming-Jie and {Zhou}, Zeng-Rong and {Wei}, Hai-Rui},
        title = "{Controlled phase gate and Grover's search algorithm on two distant NV-centers assisted by an NAMR}",
      journal = {Quantum Information Processing},
         year = 2020,
        month = may,
       volume = {19},
       number = {6},
          eid = {187},
        pages = {187},
          doi = {10.1007/s11128-020-02682-w},
       adsurl = {https://ui.adsabs.harvard.edu/abs/2020QuIP...19..187H}
}

@article{20,
  title = {Scalable superconducting qubit circuits using dressed states},
  author = {Liu, Yu-xi and Sun, C. P. and Nori, Franco},
  journal = {Phys. Rev. A},
  volume = {74},
  issue = {5},
  pages = {052321},
  numpages = {10},
  year = {2006},
  month = {Nov},
  publisher = {American Physical Society},
  doi = {10.1103/PhysRevA.74.052321},
  url = {https://link.aps.org/doi/10.1103/PhysRevA.74.052321}
}

@article{21,
   title={Coherent control of an NV−center with one adjacent13C},
   volume={16},
   ISSN={1367-2630},
   url={http://dx.doi.org/10.1088/1367-2630/16/9/093043},
   DOI={10.1088/1367-2630/16/9/093043},
   number={9},
   journal={New Journal of Physics},
   publisher={IOP Publishing},
   author={Scharfenberger, Burkhard and Munro, William J and Nemoto, Kae},
   year={2014},
   month={Sep},
   pages={093043}
}

@article{22,
   title={QuTiP: An open-source Python framework for the dynamics of open quantum systems},
   volume={183},
   ISSN={0010-4655},
   url={http://dx.doi.org/10.1016/j.cpc.2012.02.021},
   DOI={10.1016/j.cpc.2012.02.021},
   number={8},
   journal={Computer Physics Communications},
   publisher={Elsevier BV},
   author={Johansson, J.R. and Nation, P.D. and Nori, Franco},
   year={2012},
   month={Aug},
   pages={1760–1772}
}

@article{23,
title = {QuTiP 2: A Python framework for the dynamics of open quantum systems},
journal = {Computer Physics Communications},
volume = {184},
number = {4},
pages = {1234-1240},
year = {2013},
issn = {0010-4655},
doi = {https://doi.org/10.1016/j.cpc.2012.11.019},
url = {https://www.sciencedirect.com/science/article/pii/S0010465512003955},
author = {J.R. Johansson and P.D. Nation and Franco Nori}
}
\end{document}